\documentclass[fleqn,twoside]{article}
\usepackage{espcrc2}

\usepackage{graphicx}
\usepackage[figuresright]{rotating}

\newcommand{\AmS}{{\protect\the\textfont2
  A\kern-.1667em\lower.5ex\hbox{M}\kern-.125emS}}

\title{Status of resummed predictions for QCD final state observables}

\author{Mrinal Dasgupta\address{Theory Division, CERN\\ 
        CH-1211, Geneva 23, Switzerland.}}%

\begin{document}

\begin{abstract}
We provide a brief review of the current status of resummed predictions for QCD final state observables such as event shapes and jet rates in a variety of different hard processes. Particular emphasis is given to more recent developments such as the study of non global observables, development of generalized resummation formulae and resummations for new types of event shape variables.
\hfil [{\tt CERN-TH/2003-227}]
\end{abstract}

\maketitle

\section{INTRODUCTION}

All order resummed predictions have long been a common method to extend the predictive power of theoretical QCD estimates for several observables 
compared to that provided by fixed order perturbative computations alone. 
This applies in particular to observables that are measured close to their exclusive limits, such as event shape variables in $e^{+}e^{-}$ annihilation 
\cite{CTTW} where the highest statistics are typically found
close to the two jet limit. 

More generally, if a physical variable such as a typical event shape, defined for an arbitrary hard process, is constrained, by observation/measurement, to be near its Born value , it turns out to be sensitive to soft emissions. 
In fact if the deviation from the Born value of an observable (which we can take to be zero for the following discussion) is denoted by $V$, near the Born limit the cross-section  for values up to $V$ ($V \ll 1$) 
is dominated by soft and/or  collinear logarithms. In particular the $n^{\mathrm{th}}$ order perturbative estimate typically goes as :
\begin{equation}
R^{(n)}(V,Q) \sim \left ( \alpha_s(Q) \ln^2 {V} \right)^{n}+\cdots
\end{equation}
where the dots denote terms less singular in the $V \to 0$ limit and $Q$ is a scale relevant to the underlying hard process. 
Hence the predictive power of fixed order perturbation theory 
is spoilt in the small $V$ region, since the smallness of the expansion parameter $\alpha_s(Q)$ is compensated by the presence of large logarithms with upto two powers of $\ln 1/V \equiv L$ for each power of $\alpha_s$.

By now there are solid techniques in place to handle such logarithmic behaviour to all orders, which result in 
improved predictions valid over a much larger range of variable values than the fixed order computations alone, which become meaningless at very small $V$.
These resummation techniques rely mainly on factorisation methods (broadly speaking), which in turn follow from QCD coherence properties and the observables dependence on final state emissions being factorisable, usually in some appropriate transform space \cite{CTTW}. 
Due to the latter property not all variables turn out to belong to the class of resummable observables--an example is jet rates defined with the 
JADE jet algorithm \cite{JADE}.

While resummed predictions are often needed to improve perturbative accuracy, they also serve as a pathway to approach the non-perturbative domain one of the most challenging and certainly the least understood aspect of QCD.
Specifically, to access large non-perturbative effects one has to enter the low $k_t$ domain, where $k_t$ is the transverse momentum of a typical gluon emission, leading to a measured value of the observable. 
Hence in the small $k_t \sim VQ$ region, aside from large logarithms one also has power corrections $\lambda^n/(VQ)^n$ where the $\lambda$ are non-perturbative coefficients. While for $VQ \sim \Lambda_{{\mathrm{QCD}}}$ 
the problem becomes highly non-perturbative, nevertheless for hard enough processes a wide range of values exists, $\Lambda_{\mathrm{QCD}} \ll VQ \ll Q$, where both perturbative resummations and non-perturbative power corrections are important effects. Exploring this region quantitatively is important as it provides valuable information on the onset of confinement effects.

In short resummations are a useful probe of QCD dynamics, 
which comes into its own in the infrared region, 
where parton multiplication is copious and hadronisation effects are important. 
In what follows we shall confine the discussion to observables that do admit 
phase space factorisation and hence are resummable. 
We shall begin by considering the anatomy of a typical resummed answer and indicate what is the state-of--the art with regard to computation of the various pieces we will mention. We shall then discuss some significant recent developments, which contributed to the improvement of resummed predictions and consequently reflect an improved understanding of QCD dynamics: in particular the discovery of non global logarithms and the advent of generalised approaches to resummation. We shall also present some recent comparisons of such improved predictions with data.
\section{PROFILE OF A RESUMMED PREDICTION}
Consider an observable which has large logarithms in its perturbation expansion as below:
\begin{equation}
R(V) = 1+\sum_{n=1} \alpha_s^{n} \left ( \sum_{m=0}^{2n} R_{nm} \ln^{m} \frac{1}{V}+\mathcal{O}(V) \right ).
\end{equation}

One can naturally define leading logarithms (LL) as being the double logs $m=2n$ , next-to--leading logs (NLL) as those terms with $m=2n-1$ etc and in some cases this nomenclature is indeed employed \cite{burbglov}. 
Note however that in the regime where $\alpha_s L^2 \gg 1$, any 
truncation at ${\mathrm{N^{p} LL}}$ order, in this terminology, 
is no longer useful since one always has neglected terms $\alpha_s^n L^m$ with $2n-p >m>n$ which are larger than one. 
It should be pointed out that the state-of--the art resummation enables us to go to NNLL order in this notation.

However since variables that admit phase space factorisation typically exponentiate, it is possible to have a different (and in fact more commonly used) classification of logarithmic terms. Exponentiation means one can write \cite{CTTW}
\begin{equation}
R(V) = (1+C_1\alpha_s+\cdots) \Sigma(\alpha_s, L)+ D(V,\alpha_s)
\end{equation}

The first piece in brackets is simply a well behaved perturbative expansion 
(with no dependence on $V$) in $\alpha_s$. This piece treats for instance 
the mismatch 
between the full real and virtual emission and their soft limits, 
included in the form factor $\Sigma$. 

All the singular dependence on $V$ is contained in the form factor $\Sigma$.
It takes the form
\begin{equation}
\Sigma = e^{\left [ L g_1(\alpha_s L) +g_2(\alpha_s L) +\alpha_s g_3(\alpha_s L) +\cdots \right ]}.
\end{equation}

With this exponentiation it is now possible to define $g_1$ as a LL function (since it is accompanied by an extra logarithm $L$), $g_2$ as an NLL function (which is  purely single logarithmic $\alpha_s^{n} L^n$), $g_3$ as NNLL due to the extra factor of $\alpha_s$ relative to the single logarithmic $g_2$ piece. 
In this notation the state of the art is NLL resummation in that resummed predictions aim to compute up to the $g_2$ piece. This prediction is then matched to fixed order calculations. 
The `remainder' piece $D(V)$ vanishes in the $V \to 0$ limit .

Having clarified the basic notation and terminology we now turn our attention to some more recent developments. We begin the next section by providing a short discussion on non global observables.
\section{NON GLOBAL EFFECTS} 
Until very recently the resummed predictions  available in the literature 
were all made employing an independent emission formalism. In other words the matrix element for multiple gluon production was approximated by a factorised product of single soft emission contributions:
\begin{equation}
d \mathcal{P}_n \approx \prod_{i=1}^{n} 
C_F \alpha_s(k_{ti}) d\eta \frac{dk_{ti}}{\pi k_{ti}}.
\label{eq:indem}
\end{equation}
with $k_t$ the transverse momentum and $\eta$ the rapidity wrt the emitting jet axis.
 
The independent emission pattern (in one form or another) was used in conjunction with phase space factorisation to yield the form factor $\Sigma$. 
Independent emission essentially means one can assume each final state gluon to be emitted directly by the hard initiating parton as in an abelian theory. 
Subsequent branching of these gluons can in fact be neglected for several observables, apart from the contribution of such decays to the running of the coupling.

However the independent emission approximation (\ref{eq:indem})
was found to be insufficient at single logarithmic level (ie at the level of the function $g_2$) 
for several observables \cite{dassalng1}. 
In some of these cases, such as energy flow away from jets \cite{dassalng2}, single logarithms were in fact a leading effect due to absence of collinear enhancements, which meant that the function $g_1$ was absent. 
In general the observables for which independent emission breaks down at single-log level have one common feature responsible for this breakdown  -- they are sensitive to emissions only in a limited rapidity region.
For this reason such observables and corresponding logarithms 
are refered to as {\it{non global}}, while the complementary set of 
observables is refered to as global.

The origin of the problem is simple to understand and briefly explained below.
\begin{center}
\includegraphics*{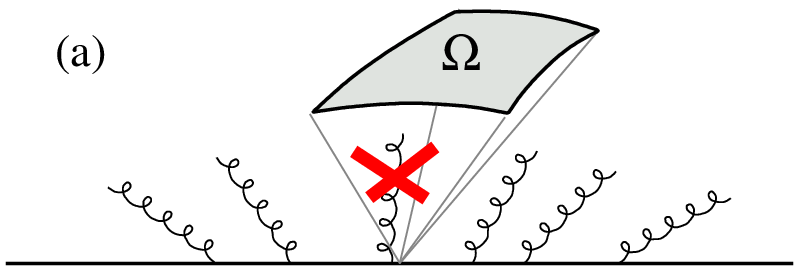}
\includegraphics*{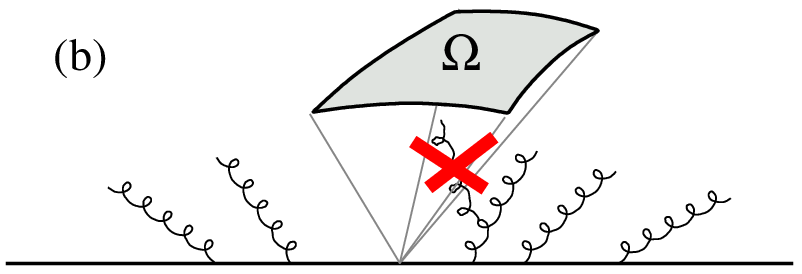}
\includegraphics*{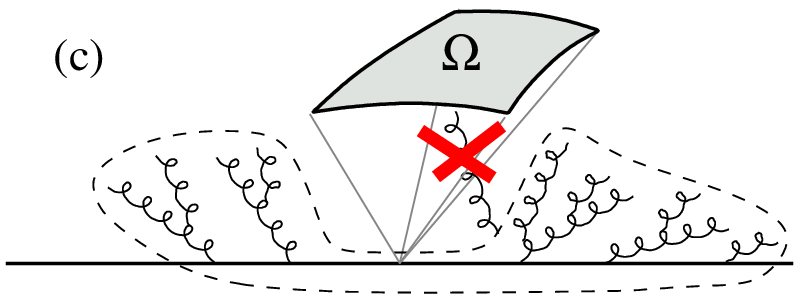}
\end{center}

Consider a measurement of say energy flow away from hard jets in a solid angle $\Omega$ as depicted above. Then by restricting, for example, the transverse energy in $\Omega$ such that $Q_\Omega \ll Q$ with $Q$ a hard scale, one is vetoing real radiation above scale $Q_\Omega$ into $\Omega$. Assuming an independent emission ansatz one needs to just veto direct emission from the hard jet lines as in diagram (a) above.

\begin{figure}[htb]
\includegraphics*{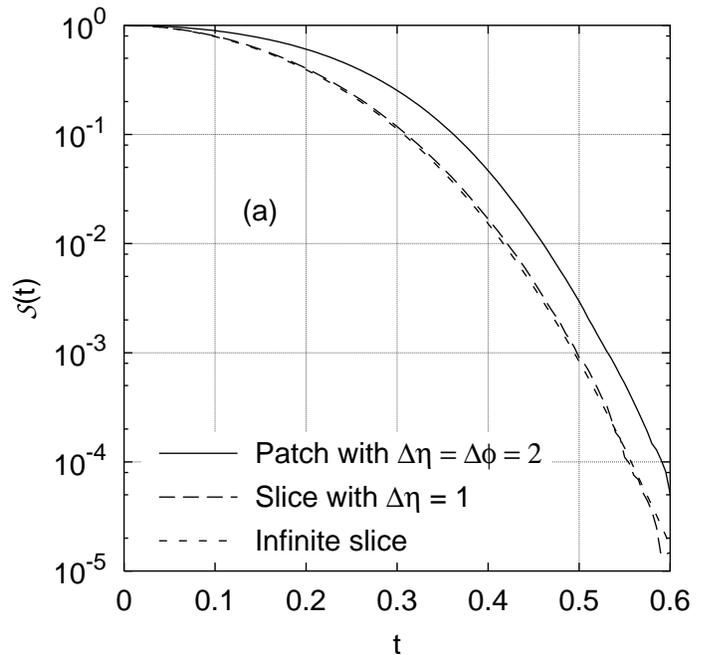}
\vspace{-1cm}
\caption{Non global function $S(t)$ plotted against $t = \int_{Q_\Omega}^{Q} \frac{dk_t}{k_t} \alpha_s (k_t) $ for different definitions of $\Omega$.}
\label{fig:ng}
\end{figure}
 This yields a single logarithmic form factor 
\begin{equation}
\label{eq:primary}
\Sigma_{\mathrm{SL}} \sim \exp[- \alpha_s \ln Q/Q_\Omega ].
\end{equation}
The above result is incorrect at single log level. This can be seen by expanding it to ${\mathcal{O}}(\alpha_s^2)$ and comparing with the logarithmic dependence of an exact fixed order estimate at ${\mathcal{O}}(\alpha_s^2)$. In fact an additional source of single logarithms is found starting 
from $\mathcal{O}(\alpha_s^2)$ which removes the discrepancy with the single logarithms in the fixed order estimate.

At order $\mathcal{O}(\alpha_s^2)$ this is just emission of a soft gluon 
with energy $\omega \sim Q_\Omega$ by a relatively harder gluon with energy $\omega'$  with $Q_\Omega \ll \omega' \ll Q$ outside $\Omega$ (see diagram (b)). 
This and analogous higher order contributions are missed by the independent emission approach which would be correct if both the gluon outside $\Omega$ and the softest gluon inside $\Omega$ were measured (vetoed). The effect of the 
softest gluon would cancel against virtual corrections at single logarithmic level. However since we are only measuring {\it{inside}} $\Omega$ the softest gluon emission is significant even in the presence of the harder emitter outside $\Omega$ and hence the gluon branching contribution appears.

The generalization of the above effect to $n^{\mathrm{th}}$ order is the coherent emission of a single softest gluon into $\Omega$ by an ensemble of $n-1$ gluons outside $\Omega$ (diagram (c) above. The $n-1$ gluon emitters are themselves ordered in energy, $\omega_0 \gg \omega_1 \gg \omega_{n-1}$ . The effect of such multiple wide-angle (non collinear-enhanced) soft 
emission is precisely a single logarithm, $\alpha_s^{n} \ln^{n} \frac{Q}{Q_\Omega}$. Resumming these terms needs a change of approach from 
the independent emission approximation. 
In Ref.~\cite{dassalng1,dassalng2}, the resummation of single logarithms 
is carried out numerically 
by using the 
dipole evolution picture which captures the essential single logarithms in the large $N_c$ limit. The resulting function is plotted in Fig.~\ref{fig:ng}.
Following this a non-linear evolution equation corresponding to the dipole evolution was derived in Ref.~\cite{BMS} and its numerical 
solution yielded identical results to those obtained in Ref.~\cite{dassalng2}.

Subsequent to the discovery of non global logs, there was some effort made at understanding how to define observables in a way such that non global effects may be minimised, which would reduce any uncertainty to do with missing non global $1/N_c^2$ suppressed effects which cannot yet be resummed. 
The reduction or removal of the non global component is also  useful for 
facilitating the phenomenology of 
multi jet event shapes and energy flows. This is because non global logs have 
thus far only been explicitly computed in the two jet case, although a similar structure is expected in the extension to multi-jet events and their calculation will also be along the same lines, employing dipole (non-linear) evolution.

In the above regard Appleby and Seymour considered the effect of jet 
clustering algorithms on non global logarithms. They found that if they defined rapidity gaps in terms of minijet energy flows (with the minijets being defined by a clustering alogorithm), rather than a sum over hadronic energies in the gap, the non global component was significantly reduced. This is essentially because the clustering algorithm has the effect of pulling 
soft hadrons (partons) out of the gap and clustering them with harder emissions outside the gap. Hence the non global contribution is not triggered except in specific 
geometrical configurations which survive the clustering, which reduces its numerical significance (see \cite{applesey} for details).

Berger, Kucs and Sterman introduced event-shape/energy flow correlations \cite{BKS} aimed at controlling non global effects in energy flow. 
This meant simultaneously restricting the energy $Q_\Omega$ in $\Omega$ alongside limiting the value of an event shape variable $V$, defined in terms of hadron momenta outside $\Omega$. 
Doing so amounts to restricting transverse momenta of soft gluons both outside the gap region, by controlling the event shape $k_t \leq VQ$, 
and inside the gap $k_t \leq Q_{\Omega}$ with $Q_{\Omega} <VQ \ll Q$. 
Non global or secondary logarithms now appear as $\alpha_s^n \ln^n \frac{VQ}{Q_\Omega}$ contributions and hence varying the values of $V$ and $Q_\Omega$ in tandem, allows one to control the significance the of the non-global contribution.

In the course of these studies the same authors also 
introduced new kinds of event shapes with an adjustable parameter $a$ that allows one to control the approach to the two 
jet limit \cite{BKS}. These variables have a parametric dependence on the transverse momenta $k_t$ and rapidity $\eta$ with respect to the thrust axis, of the form $V = \sum_i k_{ti} e^{-(1-a)|\eta_i|}$, where the sum extends over all final state partons. The resummation to NLL accutacy, for these observables can be found in Ref.~\cite{BKS}. A special case, $a=0$, of the above class of variables is the much studied thrust variable.

Dokshitzer and Marchesini \cite{dokmarch} further extended the study of event shape/energy flow correlations by showing that the non global part of the answer, which involves logarithms of $QV/Q_\Omega$,  factorises from the usual global resummation of the event shape $V$ (this time defined as a sum over all final state 
hadron contributions 
in and outside the gap).

Although some effort has been devoted to understanding and computing 
non global logarithms, the only inclusion of this effect in comparisons to data thus far has been in the case of DIS event shape variables. For details the reader is refered to Ref.~\cite{dassaldis}. A recent 
comparison of a matched resummed prediction (including non global effects and power corrections) to H1 data is shown in Fig.~2, for the thrust defined wrt the thrust axis in the Breit current hemisphere of DIS.

\section{THREE JET SHAPE VARIABLES}
An interesting and fairly recent development is that of NLL resummed predictions for three jet event shapes in $e^{+}e^{-}$ , DIS 
(two final state +one incoming jet) and hadron-hadron collisions (two incoming jets and a final state (radiated) jet+vector boson). For an example see 
\cite{BDMZ}.
\begin{figure}[htb]
\includegraphics*{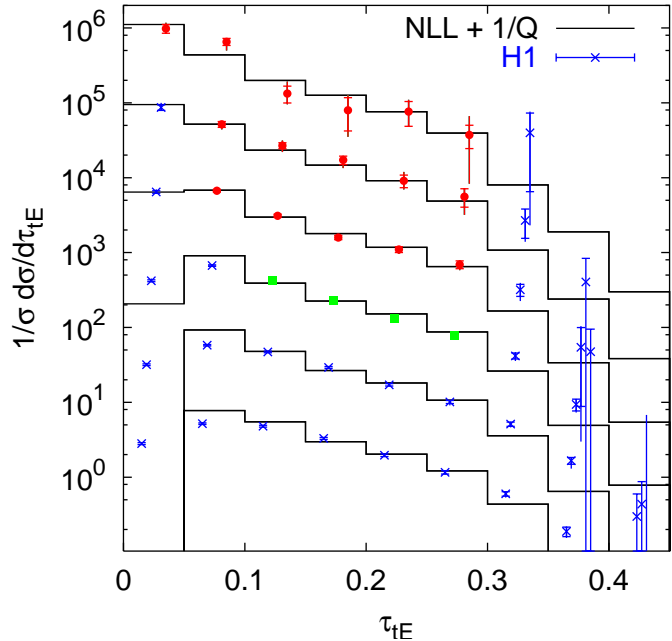}
\vspace{-1cm}
\caption{Resummed distribution for the thrust defined wrt the thrust axis in the current hemisphere of the DIS Breit frame compared to H1 data. From top to bottom represents increasing $Q$ values from 15 -- 81 GeV.}
\label{fig:DIS}
\end{figure}
For all these processes, at Born level one finds that the three hard jets lie in a plane defined say by the thrust and thrust major axes. 
Small deviations from the Born level can be accesed by studying small values of the out of plane momentum $K_{\mathrm{out}}$. 
Compared to two jet event shapes such as thrust, the three jet variables offer some new challenges. From the point of view of the resummed prediction, one as to also take account of coherent, soft interjet radiation while in the global 
two jet case only radiation emitted collinearly (effectively incoherently) by each jet contributes to NLL accuracy. The interjet radiation gives a geometry dependence at NLL level, characteristic of three jet variables. Such variables would be very interesting to study 
experimentally especially from the viewpoint of hadronisation corrections. This is a 
particularly important intermediate step along the way to understanding power behaved hadronisation effects in multijet configurations, such as in 
dijet event shapes in hadroproduction ($2 \to 2$ processes).

\section{GENERALISED FORMULAE AND AUTOMATED RESUMMATIONS}
Of late there has been significant effort and progress made in developing 
resummation formulae that are as general as currently possible and which 
reduce the need for treating different observables and different processeses  on a laborious case-by--case basis. 
The idea is to write down master formulae which have general applicability and contain process and observable dependence in a few parameters which need to be computed on a case specific basis \cite{KCLOS,BCMN,NUMSUM2}.
For example processes involving an arbitrary number of hard partons were treated in Ref.~\cite{BCMN}.

A generalised approach that aims at facilitating the computablity of resummed results , using numerical methods, and in fact automating the entire process of resummation was developed in \cite{NUMSUM2}.
The idea roughly, is to test the observable's dependence on soft and collinear emissions and determine numerically the parametric dependence on transverse momenta, rapidity and azimuth wrt the nearest leg. 
Once this dependence is obtained, the results are automatically inserted into a master formula valid for an arbitrary case 
(there however being some conditions such as globalness, that the observable has to satisfy for the master formula to be valid).
Final results are numerically obtained by means of a computer program. 
This is a particularly powerful method since almost no analytical effort is needed to generate the final results and the entire procedure is automated. 
Its advantage is particularly manifest in cases where analytical calculations are too cumbersome, requiring the inversion of multiple Mellin and/or Fourier transforms or the observable depends in a complicated way on multiple soft emissions. 
A good example of this latter point is the Durham jet finding algorithm for which the analytic computation of the full NLL 
function $g_2$ was intractable due to effects to do with recombination of soft partons. 
The numerical resummation methods  were on the other hand able to determine this function for the first time \cite{NUMSUM} .

Another example of the power of the automated resummation approach was 
provided by the computation, for the first time, to NLL accuracy 
of an event shape in hadron--hadron collisions (with a dijet final state)
-- the transverse thrust distribution \cite{NUMSUM2}. This is defined as follows:
\begin{equation}
T_\perp = {\mathrm{max}}_{\vec{n}_{\perp}} \frac{\sum_i |\vec{p}_{\perp i}.\vec{n}_\perp|}{\sum_i p_{\perp i}}
\end{equation}
where as usual the sum extends over all final state particles and $p_{\perp}$ 
is the transverse momentum wrt the beam direction while $\vec{n}$ is a unit vector found by maximising the sum in the numerator. 
The results for the different dijet production channels is illustrated in Fig.~2.

In conclusion it is worthwhile to note that there has been significant progress in the development of resummed computations in the last few years. 
The consequences of these developments are naturally of great value to experimental studies and future QCD phenomenology. Additionally they 
point towards unravelling of previously unexplored QCD dynamics and therefore to 
important progress on the theoretical front. 
\begin{figure}[htb]
\begin{center}
\hspace{-1cm}\includegraphics*{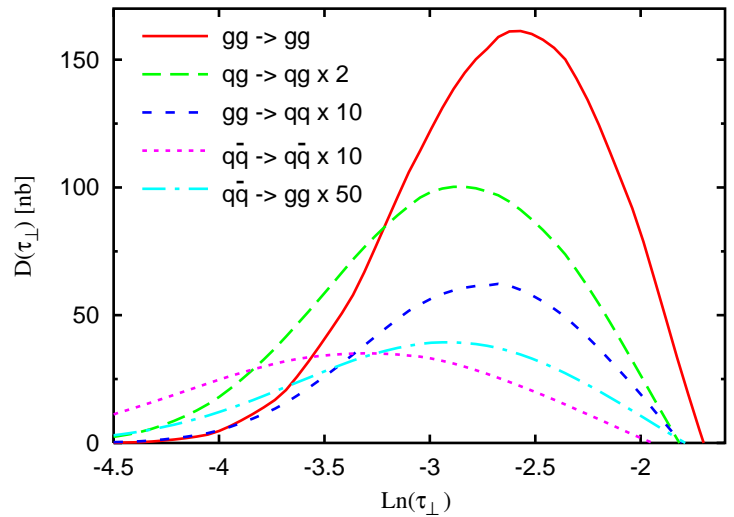}
\end{center}
\vspace{-1cm}
\caption{Plot of the transverse thrust distribution in hadronic collisions at $\sqrt{s}=1.96$ GeV.. Fig. taken from Ref.~\cite{NUMSUM2}.}
\label{fig:transthrus}
\end{figure}

\section*{ACKNOWLEDGEMENT}
I would like to thank the authors of Ref.~\cite{NUMSUM2} for the use of their figure for Fig.~3.

\end{document}